# Emergent super-antiferromagnetic correlations in monolayers of Fe$_3$O$_4$ nanoparticles throughout the superparamagnetic blocking transition

Johnathon Rackham[1], Brittni Pratt[1], Dalton Griner[1], Dallin Smith[1], Yanping Cai[1], Roger G. Harrison[2], Alex Reid[3], Jeffrey Kortright[4], Mark K. Transtrum[1], Karine Chesnel[1]

[1]Department of Physics and Astronomy, Brigham Young University, Provo, UT, 84058, USA
[2]Department of Chemistry and Biochemistry, Brigham Young University, Provo, UT, 84058, USA
[3]SIMES, Stanford Linear Accelerator, Menlo Park, CA, USA
[4]Lawrence Berkeley National Laboratory, Berkeley, CA, USA

We report nanoscale inter-particle magnetic orderings in self-assemblies of Fe$_3$O$_4$ nanoparticles (NPs), and the emergence of inter-particle antiferromagnetic (AF) ("super-antiferromagnetic") correlations near the coercive field at low temperature. The magnetic ordering is probed via x-ray resonant magnetic scattering (XRMS), with the x-ray energy tuned to the Fe-$L_3$ edge and using circular polarized light. By exploiting dichroic effects, a magnetic scattering signal is isolated from the charge scattering signal. The magnetic signal informs about nanoscale spatial orderings at various stages throughout the magnetization process and at various temperatures throughout the superparamagnetic blocking transition, for two different sizes of NPs, 5 and 11 nm, with blocking temperatures $T_B$ of 28 K and 170 K, respectively. At 300 K, while the magnetometry data essentially shows superparamagnetism and absence of hysteresis for both particle sizes, the XRMS data reveals the presence of non-zero (up to 9%) inter-particle AF couplings when the applied field is released to zero for the 11 nm NPs. These AF couplings are drastically amplified when the NPs are cooled down below $T_B$ and reach up to 12% for the 5 nm NPs and 48% for the 11 nm NPs, near the coercive point. The data suggests that the particle size affects the prevalence of the AF couplings: compared to ferromagnetic (F) couplings, the relative prevalence of AF couplings at the coercive point increases from a factor ~ 1.6 to 3.8 when the NP size increases from 5 to 11 nm.

## I. INTRODUCTION

The advance of nanotechnology heavily relies on our ability to manufacture high quality nanostructured materials and to control their behavior at the nanoscale. One type of material with high potentiality for magnetically based nanotechnologies are functional assemblies of magnetic nanoparticle (NPs). [1] When their size is below ~100 nm, magnetic NPs generally behave as individual nanomagnets, whose magnetization can be switched as a whole. A collection of such nanomagnets often present interesting magnetic properties, including superparamagnetism (SPM) [2,3] and superferromagnetism (SFM) [4,5]. These magnetic properties can be used in various applications, such as magnetic recording [6], magneto-transport [7], magneto-plasmonic [8], magneto-caloric [9], and for medical applications [10], such as contrast agents for MRI [11], hyperthermia treatment, magnetic separation, drug and gene delivery [12-17]. Due to its strong magnetization, its high chemical reactivity and its non-toxicity [18], magnetite (Fe$_3$O$_4$) is well suited for these applications. In order to optimize the functionalities of magnetite NPs in the specific applications, it is important to understand and control the magnetic response of a collection of Fe$_3$O$_4$ NPs of different sizes to various stresses, such as the application of an external magnetic field, the cooling and heating throughout the superparamagnetic blocking transition.

While the crystalline and magnetic properties of bulk Fe$_3$O$_4$ are well established [19], there is still knowledge to be gained about the magnetic behavior of collections of Fe$_3$O$_4$ nanoparticles, depending on their size, shape, concentration and type of magnetic interactions involved. At the atomic scale, the magnetization in Fe$_3$O$_4$ arises from the specific alignment of the spins carried by the Fe$^{2+}$ and Fe$^{3+}$ ions, which are distributed on a lattice of octahedral and tetrahedral sites throughout a spinel crystallographic structure [20]. Under exchange interactions, including

Zener double-exchange [21-22], the $Fe^{2+}$ and $Fe^{3+}$ spins align ferrimagnetically, thus yielding a net (non-zero) magnetic moment at the scale of the crystallites. [23] When in a nanoparticle form, and if the NP size is below about 125 nm [24], each NP is essentially monodomain (single magnetic domain extending throughout the NP), and therefore carries a massive magnetic moment, which we here call "nanospin".

When the NP size is smaller than ~ 20 nm, and if the distance between NPs is large enough so that interparticle interactions are negligible, a collection of $Fe_3O_4$ NPs typically exhibits SPM at room temperature, where each nanospin freely fluctuates independent of the neighboring nanospins but will eventually align under the application of an external magnetic field. However, there often exist non-negligeable interparticle interactions, including magnetic dipole interactions, causing the material's response to deviate from pure SPM. These interactions can lead to various interparticle orderings, such as ferromagnetic (FM), i.e., parallel alignment and antiferromagnetic (AF), i.e., antiparallel alignment of neighboring nanospins.

In this paper we specifically focus on unveiling interparticle FM and AF correlations occurring out-of-plane in Fe3O4 NPs assemblies throughout the SPM blocking transition. Because these correlations occur between nanospins (and not between atomic spins like for typical FM and AF phases), we sometimes refer to it as superferromagnetic (SFM) and super-antiferromagnetic (SAF) correlations. While the SAF term was initially introduced by Louis Néel [25] to designate the enhancement of the antiferromagnetic susceptibility in nanoparticles made of antiferromagnetic material [26], the use of SAF has not yet been standardized. Some authors used the SAF term to designate a three-state magnetic model with competing antiferromagnetic and pairing interactions in superconductors [27]. In our case, we simply suggest the use of the SAF term to designate interparticle antiferromagnetic correlations in assemblies of nanoparticles, following the same logic in which the more common term "superparamagnetic" (SPM) is used. The prefix "super" in SPM refers to the fact that the paramagnetic behavior no longer refers to individual atomic spins but to the nanospins, carried by individual NPs. Likewise, the prefix "super" in our suggested use of SAF refers to antiferromagnetic correlations between the nanospins (instead of atomic spins). We emphasize that these correlations are here not caused by any kind of exchange coupling, but instead by magnetic dipole interactions between nanospins, or with an external magnetic field when present.

The macroscopic magnetic behavior of $Fe_3O_4$ NPs and associated SPM-blocking transition has been widely studied via various techniques, such as magnetometry [28-31], Mössbauer spectrometry [32], electron paramagnetic resonance [30], and muon spin resonance [33]. To explain the observed data, in particular the relaxation dynamics of magnetic NPs in the SPM phase, sophisticated models have been developed, accounting for various magnetic anisotropies (magneto-crystalline, magnetostatic, strain, surface and shape) but also for possible interparticle interactions.[34] Additionally, Mossbauer studies have allowed to measure the effect of weak interparticle interactions on the SPM relaxation time.[35] However, this type of data is collected over a macroscopic number of NPs and does not provide spatial information about local nanoscale orderings. Limited work has been done to visualize the nanoscale magnetic correlations in assemblies of $Fe_3O_4$ NPs, using techniques such as magnetic force microscopy [36], x-ray magnetic spectroscopy [37] and neutron scattering [38], and the NP spatio-temporal behavior at the nanoscale still needs to be fully uncovered.

One important question to investigate is the nature of the magnetic correlations between individual nanospins, and its dependence on magnetic field. If the material is in the SPM phase, the net macroscopic magnetization $M$ of a NP assembly vanishes ($M = 0$) when the external magnetic field is released to zero ($H = 0$), due to spatial magnetic randomness, combined with nanospin fluctuation over time. At the macroscopic scale $M = 0$, but at the microscopic scale each NP still carries an intrinsic nanospin at any given time. An interesting question to explore is what kind of arrangements a collection of nanospins may generally adopt to yield $M = 0$. At any given time (setting magnetic fluctuations aside) both a random distribution of nanospins and an AF arrangement of nanospins can yield $M = 0$. Additionally, one must account for possible fluctuations. While in the blocked phase, the magnetic order may remain static over a certain period of time, nanospin fluctuations will eventually lead to M = 0 at higher temperature in the SPM phase. In our 2D $Fe_3O_4$ NP assemblies, do inter-particle magnetic correlations exist? If so, what is the nature of these correlations, and how do they depend on field, temperature and particle size?

To answer these questions, one needs to be able to probe NP materials at the nanoscale, and only a few techniques allow this to be done, including polarized electron microscopy, neutron scattering and x-ray scattering. Small angle neutron scattering (SANS) [39] has traditionally been the method of choice to probe nanoscale magnetic correlations in magnetic materials. SANS studies have been carried on bulk $Fe_3O_4$ NPs [40]. However, there are a few limitations associated with neutron scattering techniques, in particular the need of massive samples to produce sufficient scattering signal, making data collection on thin monolayers of NPs challenging. Electron microscopy, on the other hand, allows the probing of very thin materials locally, offering sub-nanometric resolution. When polarized electron microscopy is available, it yields a magnetic contrast allowing one to identify the magnetic state of individual NPs [41]. However, polarized electron microscopy is challenging to implement and necessitates extremely thin materials that do not exceed a few nanometers, making the imaging of bigger NPs challenging.

To probe magnetic correlations in monolayers of $Fe_3O_4$ NPs, we utilize here an original synchrotron x-ray-based technique, x-ray resonant magnetic scattering (XRMS) [42], which allows one to surpass the mentioned limitations encountered with using neutrons or electrons. The use of synchrotron light enables the tunability of the x-ray energy to resonance edges of the magnetic elements contained in the material, here the Fe–$L_3$ edge, providing element selectivity and magneto-optical contrast. At the Fe–$L_3$ edge energy ~708 eV (soft x-ray range), the wavelength of the light is ~1.75 nm, which is perfectly suited for accessing spatial correlations in the range of a few nanometers. Additionally, strong incident intensity and resonant scattering effects allow for rapid data collection from NP monolayers. Finally, the XRMS technique allows to visualize magnetic correlations even when the net magnetization of the material is zero, which generally happens in SPM NPs when the magnetic field is released. XRMS is therefore a method of choice to study nanoscale magnetic correlations in NP monolayers.

Very few XRMS studies of magnetic NPs have been reported so far. To our knowledge, our study [43] was the first to report XRMS data collected monolayers of $Fe_3O_4$ NPs, along with modeling [44]. Previous studies on Co NP assemblies [45] have demonstrated the possibility to access inter-particle magnetic correlation via the use of XRMS in linear polarization, by monitoring the variation of the XRMS signal with an external magnetic field. Here we show a complementary approach, by using XRMS in circular polarization and by exploiting dichroic effects to extract information on the charge correlations and on the magnetic correlations, separately. We use this circularly polarized XRMS method to extract inter-particle magnetic correlations on different sizes of $Fe_3O_4$ NPs, 5 and 11 nm, throughout the magnetization process, at different temperatures, above and below the SPM blocking transition.

## II. METHODS

Below we provide details about the experimental steps and methods employed for this study, including the NP fabrication procedure, the transmission electron miscroscopy (TEM) imaging, the vibrating sample magnetometry (VSM) measurements and the XRMS measurements using synchrotron radiation.

### A. Nanoparticle fabrication

Our $Fe_3O_4$ NPs were prepared at Brigham Young University via an organic solution method based on the thermal decomposition of an iron precursor in the presence of oleic acid. The use of oleic acid yields the formation of a ligand shell around each particle, which prevents particle clustering and leads to relatively uniform NP size. The particle size is tuned by adjusting various parameters in the preparation. Two particle sizes are considered in this study: 5 nm (M5) and 11 nm (M11). The 5 nm particles were prepared via the thermal decomposition of Fe(acac)$_3$ in the presence of phenyl ether, hexadecane, oleylamine and oleic acid at 275 ºC for 30 min [46,47]. The 11 nm particles were prepared by heating an Fe(III) oleate in the presence of oleic acid and octadecene at 320 ºC for 30 min [48]. After cooling to room temperature (25 ºC), the NPs were precipitated in ethanol, and separated by centrifugation at a speed of 5000 rpms for 15 min. After successive rounds of precipitation, a black powder was obtained, which was then used for the TEM, VSM and XRMS measurements shown in this paper. Additionally, we have carried x-ray diffraction (XRD) measurements, already reported in [30], for which Rietveld refinements confirmed a nearly pure (>99%) magnetite Fe$_3$O$_4$ crystallographic phase. The XRD data also showed that the NPs are all monocrystalline.

*B. TEM imaging*

To study their shape and size, the NPs were imaged via TEM. The TEM images were collected at BYU on a ThermoFisher Scientific Tecnai F20 UT instrument operating at 200 kV. For the TEM imaging, the NPs were deposited either on thin carbon membrane grids or silicon nitride (Si$_3$N$_4$) membranes. To prepare these TEM membranes, the prepared NPs, once in a powder form, were dissolved in either chloroform or toluene, and a drop of the solution was dropped onto the membranes, where the solvent would rapidly evaporate, leaving a self-assembly of NPs. The concentration of the solution was incrementally adjusted in order to achieve a monolayer of NPs - too concentrated solutions yielding multilayer stacking while not enough concentrated solutions yielding sparse islands of NPs.

*C. Magnetization and FC/ZFC and data*

The magnetization data was collected at BYU via VSM on a Quantum Design Physical Properties Measurement System (PPMS) that includes a 9 T superconducting magnet and a cryogenic sample holder using liquid helium allowing to cool samples down to below 10 K. For the VSM measurements, the NPs were inserted in a powder form into a capsule and compacted into a cylindrical pellet of about 2 mm in diameter and 1 mm in thickness. The capsule was securely mounted on the VSM holder and tightened with quartz pieces to prevent internal motion while the holder is vibrating inside the detection coil. Magnetization loops were collected at various temperatures using a continuous sweeping of the superconducting magnet at about 50 Oe/s. Zero-Field Cooling (ZFC) and Field Cooling (FC) data were collected by typically warming from 10 K up to 400 K at a speed of 1 K /s under a magnetic field of 100 Oe.

Powders of NPs were used for the magnetometry measurements because thin monolayers of NPs would not provide measurable VSM signal (and also the membranes would break under the vibration). It is possible that the magnetic response of powders of NPs measured via VSM could slightly differ from the magnetic response of monolayers of the same NPs. However, as described in section III, the TEM data shows that the NPs in the assemblies tend to close pack, so it is reasonable to assume that the average interparticle distance and the magnetic interactions are similar in the powders and the films, for a given NP size. Also, the TEM images show that the NPs are nearly spherical so one can assume no significant shape magnetic anisotropy, and therefore the magnetic response should not significantly depend on the direction of the applied magnetic field. We note, however, that the magnetic field in the VSM measurement was applied out-of-plane with respect to the pellet, and the magnetic field in the XRMS measurement (see below) was also applied out-of-plane with respect to the NP monolayers.

*D. X-ray resonant magnetic scattering (XRMS)*

The XRMS data was collected on a coherent diffractive imaging instrument located at Beamline 13, at the Stanford Synchrotron Radiation Light Source (SSRL) synchrotron facility at Stanford Linear Accelerator Laboratory (SLAC). The x-ray light, produced by an Elliptical Polarization Undulator (EPU) was circularly polarized with a degree of polarization close to 99%. Deposited on thin Si$_3$N$_4$ membranes, the NP monolayers were probed in transmission geometry, with the x-ray beam at normal incidence, and 2D scattering patterns were measured downstream on a CCD camera, as illustrated in our previous publication [43]. The 1" square CCD detector included 2048×2048 pixels of 13× 13 µm in size and was positioned at 10 cm downstream of the sample. The experiment was carried out in a vacuum scattering chamber to limit absorption of the soft x-rays by air. The vacuum chamber included an electromagnet, allowing one to apply an in-situ magnetic field from −3000 Oe up to +3000 Oe, in the direction of the incident x-ray, i.e., out-of-plane with respect to the membranes. In that geometry, the XRMS probe is essentially sensitive to the out-of-plane component of the magnetization with respect to the plane of the layer. The instrument also included a cryogenic sample holder using liquid helium, allowing to cool the samples down to 15 K. To optimize the magneto-optical contrast, the x-ray energy was finely tuned to a magnetic resonance at 706 eV, located within the $L_3$ edge of Fe. To find the exact energy of the magnetic resonance, prior x-ray magnetic circular dichroism (XMCD) data was collected under a magnetic field of ~ 6000 Oe [37] showing a triple peak feature characteristic of $Fe_3O_4$. For the XRMS measurement, the x-ray energy was set to the first peak at 706 eV. In order to separate the magnetic scattering signal from the charge scattering signal, the XRMS data was collected at opposite helicities of circular polarization. The exposure and collection time combined was around 20 sec per CCD image. To increase the signal to noise, typically 10 of these images were successively collected, for a total measurement time of ~200 sec at a given temperature, field and polarization, and averaged to extract the XRMS signal presented in this paper.

*E. Data fitting*

The data was fit using the python library lmfit, a useful extension of the scipy optimization methods, for non-linear least-squares minimization to fit the charge intensity and magnetic ratio curves [49]. We used the implementation of the Levenberg-Marquardt algorithm found in lmfit for the residual optimization. Further discussion of the model used for the fitting process can be found in section V.

III. MATERIAL PROPERTIES

Here we describe the basic structural and magnetic properties of the studied NPs. Information about the particle shape, size and tendency to self-assemble is obtained from TEM images. Information about the SPM blocking transition and associated blocking temperature is extracted from ZFC/FC curves. Finally, the occurrence of hysteresis is discussed based on magnetization data.

*A. Structural properties*

The TEM image of M5 in Fig.1a indicates that the 5 nm NPs are typically spherical and relatively uniform in size. A statistical analysis of a collection of NPs on the image yields an average particle size of 5.3 ± 0.7 nm. The TEM image shows that the NPs tend to self-assemble in a closely packed hexagonal lattice, forming a uniform monolayer.

The TEM image of M11 in Fig.1b shows that the 11 nm NPs are also spherical but less uniform in size. The statistical analysis yields an average particle size of 11.3 ± 2.5 nm. These NPs also tend to self-assemble in a hexagonal manner, but due to the wider size distribution, the extent of the lattice is spatially limited, leading to sparse islands with different orientations. The sharp change in contrast visible within some NPs suggests occasional crystal twinning. The magnetization and XRMS data however indicate that the NPs essentially remain magnetically monodomain.

*B. Superparamagnetic blocking transition*

The ZFC/FC data collected at 100 Oe on M5 in Fig.1c shows a very sharp transition between the SPM and blocked states, with a blocking temperature $T_B$ = 28 K. The sharpness of the transition reflects the narrow size distribution.

The ZFC/FC data collected at 100 Oe on M11 in Fig.1d shows a broader transition with an average blocking temperature $T_B$ ~ 170 K. The drastic increase in $T_B$ is due to the increase in NP size from 5 to 11 nm. The peak broadening reflects the wider size distribution.

*C. Magnetization data*

Magnetization loops were measured at 300 K (above $T_B$) and at 20 K (below $T_B$). These points in temperature are indicated by green dots on the ZFC/FC curves. The magnetization loops for M5 in Fig.1e show a smooth S-shape, which can be modeled by the Langevin function, characteristic of SPM. The absence of hysteresis confirms a pure SPM phase for the 5 nm particles. For M11, the magnetization loops shown in Fig.1f reveal very little hysteresis at 300 K but some significant hysteresis, up to $H_c$ = 300 Oe at 20 K, well below $T_B$. Magnetic hysteresis can appear in the blocked state when the magnetic NPs exhibit internal magnetic anisotropy. [35] However, the TEM micrographs showed that our NPs are mostly spherical and therefore do not exhibit significant magnetic shape anisotropy. On the other hand, our XRMS data clearly show the presence of interparticle magnetic correlations. These observations suggest that the magnetic hysteresis may here be caused not solely by magnetic anisotropies but also by inter-particle magnetic couplings, as confirmed by the XRMS data presented below.

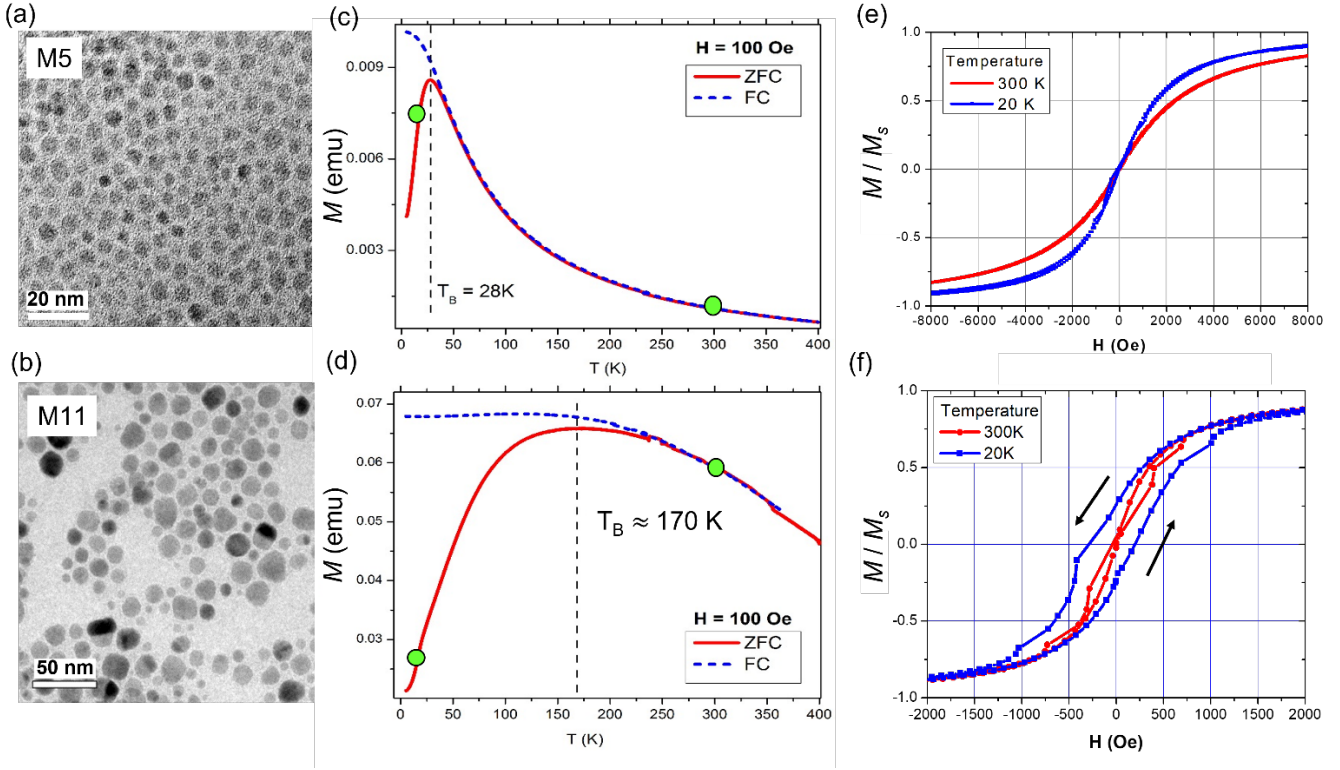

FIG.1. Structural and magnetic properties of the 5 nm NPs (M5) and 11 nm NPs (M11). (a) TEM view of M5. (b) TEM view of M11. (c-d) FC/ZFC data measured via VSM under a field $H$ = 100 Oe on (c) M5 and (d) M11. (e-f) Magnetization loops measured via VSM at 300 K and 20 K on (e) M5 and (f) M11, respectively. The temperature points for magnetization data are located by green dots on the FC/ZFC curves. The TEM images were collected on thin monolayers of NPs whereas the VSM data was collected on powders made of same NPs.

## IV. X-RAY MAGNETIC SCATTERING RESULTS

To investigate the nanoscale magnetic orderings between the NPs, we collected XRMS data on monolayers of the various $Fe_3O_4$ NPs. In our previous work [43], we showed how the x-ray energy is first tuned to the Fe-$L_3$ edge using x-ray magnetic circular dichroism (XMCD) and demonstrated how to exploit the polarization of the light to extract a magnetic signal from the scattering patterns. For this purpose, we introduced a magnetic ratio $R_M$, aimed to isolate the magnetic signal from the charge signal in the scattering patterns. Here we apply this approach to probe magnetic orderings that may occur with the out-of-plane component of the nanospins and compare it to the charge ordering within the NP monolayer, throughout the SPM blocking transition. First, we show 2D scattering patterns and their associated 1D scattering profiles. We then review the steps for separating the magnetic signal from the charge signal in these 1D profiles. Finally, we look at the dependence on magnetic field and on temperature for the 5 and 11 nm NPs.

### A. Charge and magnetic scattering signal separation

The XRMS patterns collected on M5 (Fig.2a) and on M11 (Fig.2b) inform on the lattice formed in the respective NP assemblies. Both patterns have the shape of a ring, here in the momentum space. The ring radius is inversely proportional to the average inter-particle distance in the NP assembly. The radius is larger for M5 than for M11, consistent with a smaller inter-particle distance for the 5 nm NPs. Also, the ring is relatively narrower for M5 reflecting a more highly ordered NP lattice than for the 11 nm NPs.

In order to extract quantitative information on the charge structure and the magnetic structure in the NP assemblies, we first integrated the 2D scattering patterns azimuthally so to generate 1D datasets of the scattering intensity $I(q)$ as a function of the momentum $q$. Next, we separated the charge scattering from magnetic scattering signal by exploiting the light polarization as follows. As derived in [50], the scattering intensity may be written in terms of scattering amplitudes:

$$I_\pm = |A_c \pm A_m|^2$$
$$= |A_c|^2 \pm (A_c A_m^* + A_m A_c^*) + |A_m|^2$$

where $I_+$ and $I_-$ represent the scattering intensity collected with positive and negative circular polarization, respectively; $A_c$ and $A_m$ represent the amplitudes of the charge and magnetic scattering, respectively. By construction, $A_c$ and $A_m$ are complex quantities, and the * symbol denotes the complex conjugate. To separate $A_c$ and $A_m$, at a first order, the following approach is adopted. On one hand, we consider the average of intensities between opposite helicities of the circular polarization (which also corresponds to the intensity $I_0$ one would measure in linear polarization):

$$I_0 = (I_+ + I_-)/2 = |A_c|^2 + |A_m|^2.$$

In our specific data set, however, $|A_m|^2$ turns out to be negligible compared to $|A_c|^2$. Consequently:

$$I_0 = (I_+ + I_-)/2 \approx |A_c|^2.$$

This approach differs from another approach adopted in other studies [45] where the sum $I_+ + I_-$ is compared at different field values $H$, typically at saturation and at remanence ($H = 0$) and the difference give access to the magnetic part $|A_m|^2$. In our case, the sum $(I_+ + I_-)(q)$ shows visibly no dependence with $H$, as demonstrated in Fig.2c and Fig.2d for M5 and M11, respectively, confirming that $|A_m|^2 \ll |A_c|^2$ here. Additionally, the fact that $|A_c|^2$ is essentially not changing when $H$ is varied indicates that the NPs, sitting solidly on the membrane, are not physically moving when applying a magnetic field up to 3000 Oe.

On the other hand, we consider the difference of the scattering intensities measured at opposite helicities:

$$I_+ - I_- = 2(A_c A_m^* + A_m A_c^*) = 2Re(A_m A_c^*)$$

This difference, a real number, may be either positive or negative in sign, and will typically switch sign when the internal magnetization of the material is reversed via the reversal of the applied magnetic field, as demonstrated by our previous studies [43]. As it is conventionally done in XMCD, a dimensionless dichroic ratio may be derived from the difference:

$$R_D = \frac{I_+ - I_-}{I_+ + I_-} = \frac{2Re(A_m A_c^*)}{2(|A_c|^2 + |A_m|^2)} \approx \frac{Re(A_m A_c^*)}{|A_c|^2}$$

However, $R_D$ depends on both $A_c$ and $A_m$ amplitudes. Our approach is to instead consider the following "magnetic ratio" $R_M$:

$$R_M = \frac{I_+ - I_-}{\sqrt{I_+ + I_-}} = \frac{\sqrt{2}\,Re(A_m A_c^*)}{\sqrt{|A_c|^2 + |A_m|^2}} \approx \frac{\sqrt{2}\,Re(A_m A_c^*)}{|A_c|}.$$

Now, our data suggests that there is no significant dephasing between $A_c$ and $A_m$, so here $Re(A_m A_c^*) \approx |A_c||A_m|$ and consequently $R_M$ does not much depend on the charge scattering amplitude $|A_c|$ but essentially relies on the magnetic scattering amplitude:

$$R_M \propto |A_m|.$$

That way, the magnetic signal can be separated from the charge signal to a first approximation. Incidentally, $R_M$ does not only depend on the magnitude of $A_m$ but also, via the phase factor, does switch sign when the magnetization of the material is reversed. In order to compare the charge scattering and magnetic scattering signals with having same dimension, we chose to consider the square of $R_M$:

$$R_M^2 \propto |A_m|^2$$

which, like $I_0$, has the dimension of an intensity.

In summary, we focus here on two independent quantities that provide us with information on the charge structure and the magnetic structure in the material:

$$I_0 \approx |A_c|^2(q)$$

$$R_M{}^2 \approx |A_m|^2(q).$$

As an illustration, $R_M{}^2$ is represented next to its associated $I_0$ signal in Fig.2e and Fig.2f, for M5 and for M11, respectively. This data is measured at 300 K with $H$ varying from 3000 Oe down to zero. For both M5 and M11, $R_M{}^2$ drastically varies with $H$. Not only the magnitude of $R_M{}^2$ gradually decreases all the way down to near zero when $H$ is decreased to zero, confirming the magnetic nature of $R_M{}^2$, but its shape (dependence on $q$) changes too, indicating changes in magnetic density within the material.

The dependence of $R_M{}^2$ on the momentum $q$ contains spatial information on the nanoscale magnetic ordering. In this paper, we study how the shape of $R_M{}^2(q)$ varies with $H$ throughout the magnetization process and with temperature $T$ when cooling down below the SPM blocking transition, for each of the two particle sizes. For comparison purposes, we show, in the subsequent figures, a normalized version of $|R_M|(q)$ where the magnitude of the main peak (located at $q^*$) is normalized to 1. Comparing the shape of $|R_M|(q)$ allows to identify the emergence of various magnetic orders, such as FM and AF orders. We later quantify the amount of respective FM and AF orders, but rather than using the absolute magnitude of $|R_M|(q)$ which we found unreliable (due to various experimental instabilities) we used the relative magnitudes of the FM and AF signals, combined with magnetization data.

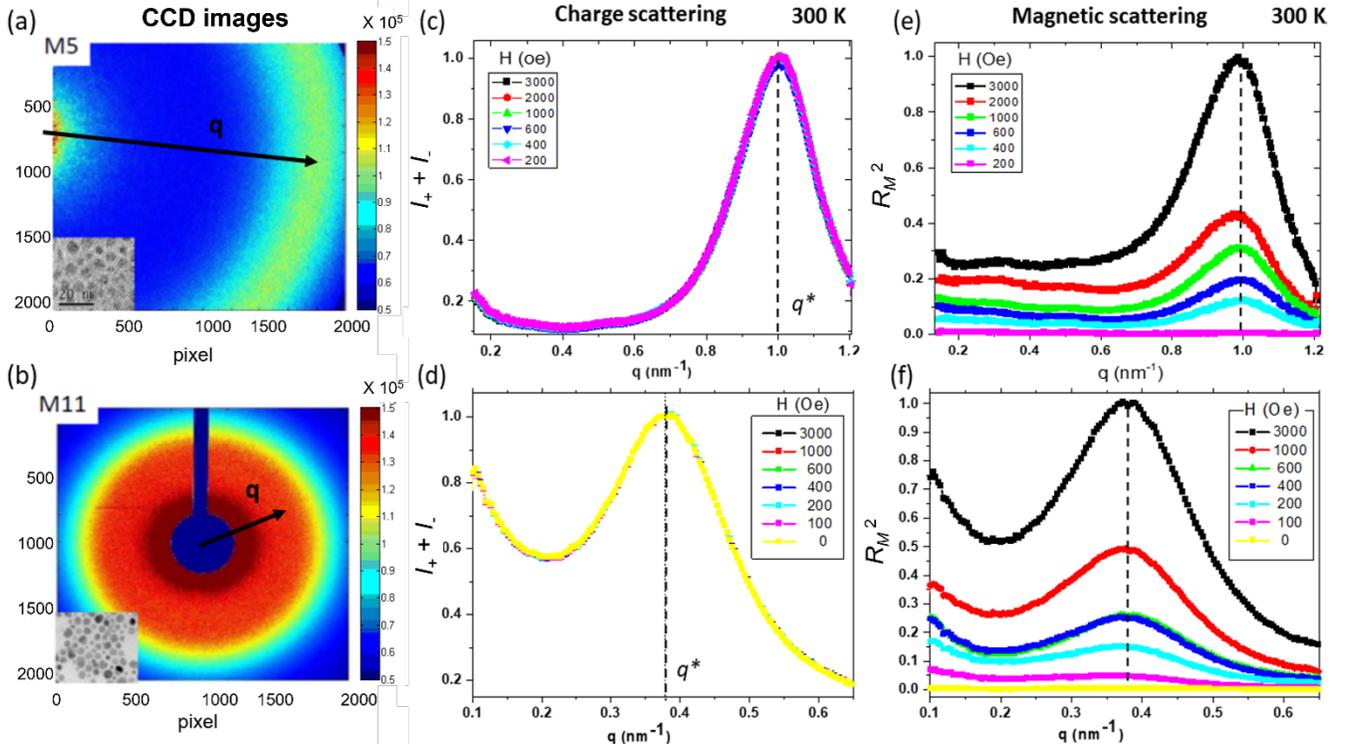

FIG.2. 2D X-ray resonant magnetic scattering (XRMS) patterns measured for (a) M5 and (b) M11 at the Fe-$L_3$ edge. (c-f) Extracted charge and magnetic scattering profiles after azimuthal integration of the 2D patterns and combination of intensities $I_+(q)$ and $I_-(q)$ measured in opposite helicities of the circular light, shown at various applied magnetic field values $H$, from 3000 Oe down to zero. The charge scattering signal is $I_+ + I_-$, with peak at $q^*$ normalized to 1, in (c) for M5 and in (d) for M11. The magnetic scattering signal is $R_M{}^2 = (I_+ - I_-)^2/(I_+ + I_-)$ with peak at $q^*$ normalized to 1 for the highest field value $H$ = 3000 Oe, in (e) for M5 and in (f) for M11.

## B. Results for 5 nm particles

The charge scattering signal plotted in Fig. 2c shows a peak at $q^* \approx 1.0$ nm$^{-1}$, suggesting an average inter-particle distance $p_0 = 2\pi/q^* \approx 6.3$ nm for M5. Given the particle average size of ~ 5.3 nm, this leaves an average gap between particles of about 1 nm, which is about the size of the oleic acid ligand shell. This confirms that the 5 nm NPs are mostly close-packed, forming a hexagonal lattice, as suggested by the TEM images on smaller fields of view.

The magnetic scattering signal at 300 K plotted in Fig.2e shows a peak located at the same location $q^* \approx 1$ nm$^{-1}$. This indicates the presence of magnetic correlations in the NP assembly, namely a FM ordering of the nanospins, where the magnetic period matches the structural one: $p_{FM} = p_0$. In this configuration, most of the nanospins point in the direction set by the external field. The FM signal is the strongest at the highest field value $H = 3000$ Oe and progressively decreases when $H$ is decreased, as the internal net magnetization $M$ also decreases.

Fig.3 shows the evolution of normalized $|R_M|(q)$ with magnetic field for M5, at 300 K above $T_B$ and at 15 K below $T_B$. On the left side, a zoomed-in view of the magnetization curve $M(H)$ in the range of [-3000 Oe, +3000 Oe] is plotted, showing the various points in field $H$ where XRMS data was collected. Next to each point, the associated degree of magnetization $M$, normalized to magnetization at saturation $Ms$, is indicated (in %). The $|R_M|(q)$ data at 300 K in Fig.3b shows a single FM peak at $q^*$ for most field values $H$ from 3000 Oe down to 400 Oe. Only when $H \leq 200$ Oe, an additional peak around $q^*/2$ emerges, suggesting the presence of some linear AF ordering of nanospins, for which the periodicity is $p_{AF} = 2p_0$. This AF peak is however much weaker than the FM peak, seemingly less than half of it. Given that at $H = 200$ Oe, the net magnetization $M$ is only 6% of the magnetization at saturation $M_s$, the amount of FM correlations covers only around 6%, and therefore the amount of AF correlations is relatively small, seemingly less than 3% (the data fitting described in the next sections consistently shows that this quantity is closer to 2.3%).

The behavior at T = 15 K, below $T_B$, is similar to that at 300 K, except that the AF peak emerges a bit earlier, at $H = 400$ Oe, when $H$ is decreased and becomes relatively stronger compared to the FM peak when $H$ approaches zero. No hysteresis was observed in the magnetization curve $M(H)$, so a symmetrical behavior is expected in respect to switching the sign of $H$.

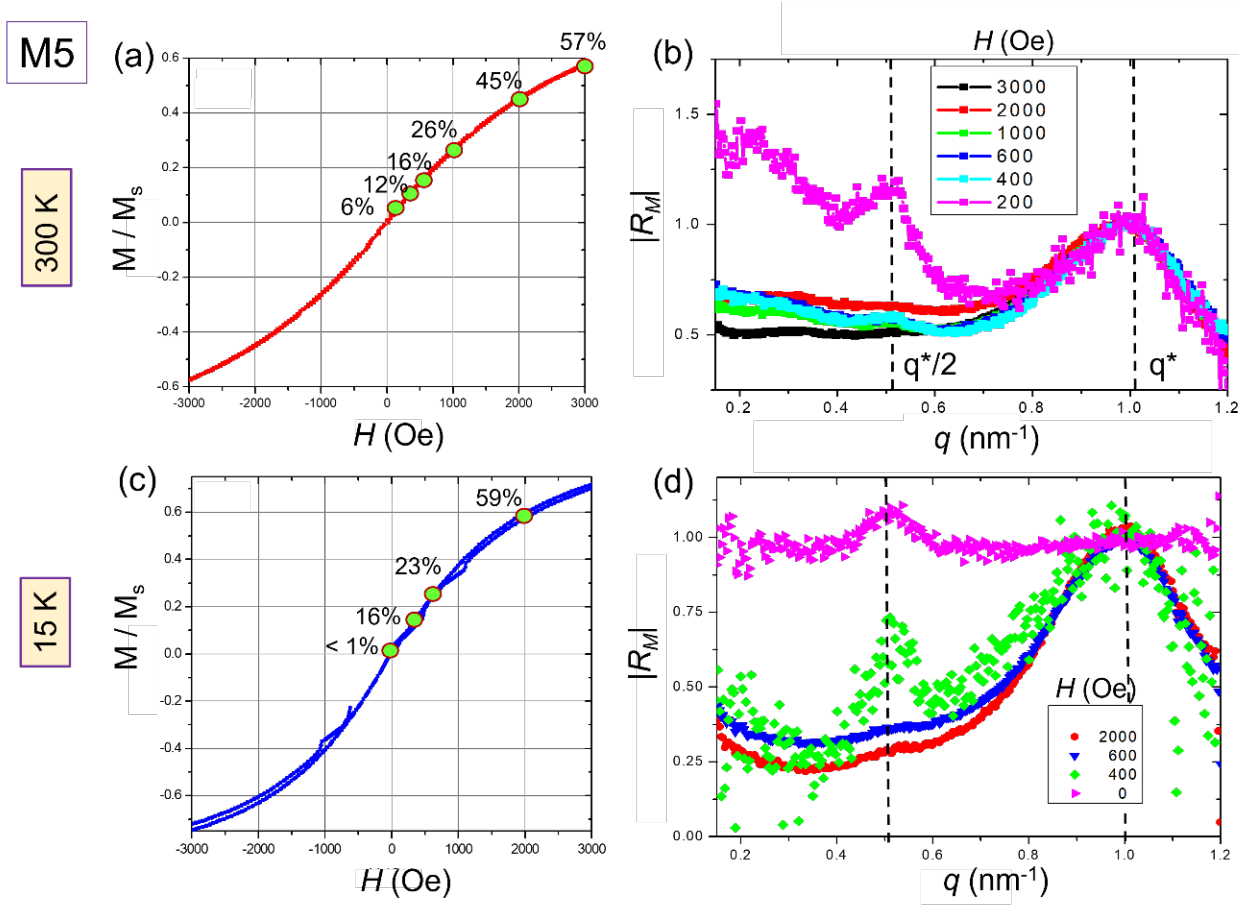

FIG.3. XRMS results for M5 at 300 K and 15 K. (a) Normalized magnetization curve $M(H)$ measured at 300 K in the ±3000 Oe range. (b) $|R_M|(q)$ data measured at 300 K at the various $H$ values indicated on the $M(H)$ curve (a). (c) Normalized magnetization curve $M(H)$ measured at 15 K in the ±3000 Oe range. (d) $|R_M|(q)$ data measured at 15 K, at the various $H$ values indicated on the $M(H)$ curve (c). For (b) and (d), the $|R_M|(q)$ data is normalized so to set the value of the peak at $q^*$ to 1.

## C. Results for 11 nm particles

The charge scattering signal for M11, plotted in Fig. 2d, shows a peak at $q^* \approx 0.38$ nm$^{-1}$, which corresponds to an average inter-particle distance of about 16.5 nm. Given the particle average size of 11.3 nm, this leaves an average gap between particles of about 5.2 nm. This loose gap accounts for the wide distribution of NP sizes, which prevents the NPs to closely pack on large areas, but rather allows the NPs to form sparse islands, as seen on the TEM image Fig.1b.

The normalized magnetic ratio signal $|R_M|(q)$ for M11 is shown in Fig.4, both at 300 K above $T_B$ (Fig.4a,b) and at 20 K well below $T_B$ (Fig.2c,d). In contrast with M5, the data for M11 generally shows a more pronounced dependence of the shape of $R_M^2(q)$ on the magnetic field $H$. The $|R_M|(q)$ data in Fig.4b collected at 300 K and $H = 3000$ Oe, shows essentially one single peak, located at same $q^*$ than the peak in the charge scattering signal in Fig.2d, corresponding to FM ordering. When the field $H$ is lowered, the FM peak remains present, but the magnitude of the signal at around $q^*/2$ progressively increases, suggesting the gradual emergence of some AF correlations in the NP assembly. The AF signal remains however limited, dominated by the FM signal.

At 20 K, on the other hand, the progression is significantly more drastic. While at $H = 3000$ Oe, the system shows mostly FM ordering with a main peak at $q^*$, the shape of $|R_M|(q)$ however drastically changes when the field $H$ is lowered, an AF signal extended on a wide $q$ range near $q^*/2$ arises and eventually surpasses the FM signal. At 20 K, M11 shows hysteresis, so at $H = 0$, there is still a strong remanence (M/Ms $\approx 25\%$).

When further decreasing $H$ down to negative values, the AF signal then strikingly increases, far dominating the FM signal, and reaches a maximum value at the coercive point $H = -Hc = -300$ Oe.

The data on M5 and M11 undeniably shows the emergence of some AF ordering of out-of-plane nanospins, when the NP assembly is cooled down below $T_B$ and when $H \sim 0$. It also shows that the amount of AF correlations with respect to the amount FM correlations depends on particle size. The data here suggests that the smaller, 5 nm NPs tend to orient mostly randomly when the field is released to zero, whereas the bigger, 11 nm particles show significant AF correlations, that are the strongest when the field is brought to the coercive point.

To solidify these qualitative observations, we fitted the data using an empirical model, which provides quantitative information about the various magnetic orders in the NP assemblies, including their relative amounts and their associated periodicities and correlation lengths.

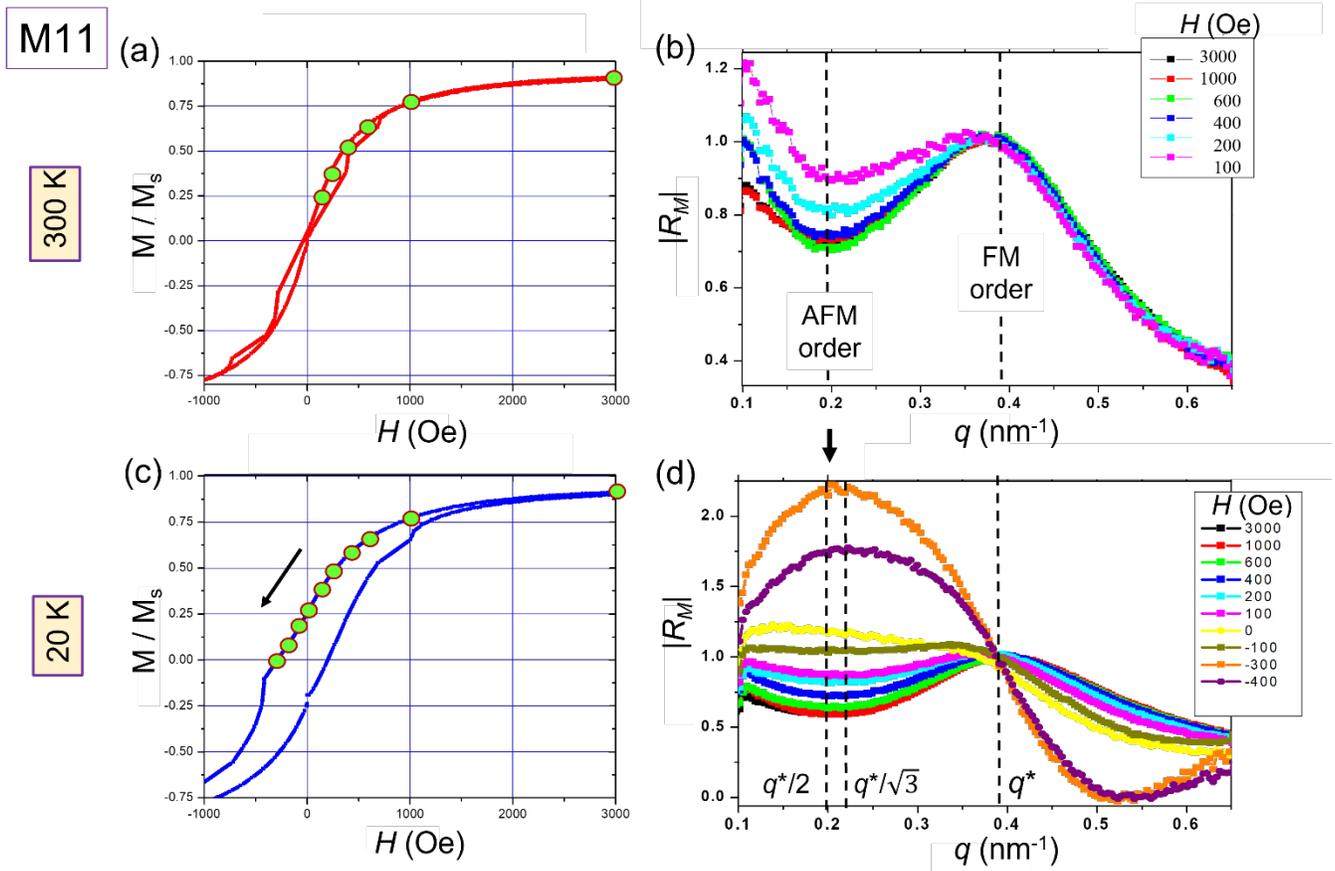

FIG.4. XRMS results for M11 at 300 K and 20 K. (a) Normalized magnetization curve $M(H)$ measured at 300 K, focusing on the positive branch. (b) $|R_M|(q)$ data measured at 300 K at the various $H$ values indicated on the $M(H)$ curve (a). (c) Normalized magnetization curve $M(H)$ measured at 20 K showing hysteresis, with coercive field $|H_c| \approx 300$ Oe. (d) $|R_M|(q)$ data measured at 15 K, at the various $H$ values indicated on the $M(H)$ curve (c) from +3000 Oe down to -400 Oe, past the coercive point. For (b) and (d), the $|R_M|(q)$ data is normalized so to set the value of the peak at $q^*$ to 1.

## V. DATA FITTING

Here we describe the empirical model we utilized to fit both the charge and the magnetic components of the x-ray scattering data. The primary goal of this fitting work is to obtain quantitative information on the various magnetic ordering components present in the material and their dependence with field and temperature for each particle size. Below, we review the model used for the fit, its mathematical formulation and its associated parameters. We then describe how the fit was conducted and optimized as a global fit on the charge and magnetic components at different field values.

*A. Model*

We adopt here an empirical model consisting in fitting the scattering data directly in the $q$ space. Best results were obtained with Lorentzian functions for fitting the charge scattering signal $I_0$, and with Gaussian functions for fitting the magnetic scattering signal $|R_M|$.

First, we fit the charge scattering signal, $I_0 = (I_+ + I_-)/2$, which generally exhibits one main peak, sitting over a diffuse scattering background, using a combination of Lorentzian functions:

$$I_0(\boldsymbol{b}; q) = b_1 \frac{b_2}{q^2+b_2^2} + b_3 \frac{b_5}{(q-b_4)^2+b_5^2} + b_6$$

where $\boldsymbol{b} = (b_1, b_2, \ldots b_6)$ represents the fitted parameters. The parameters $b_1$ and $b_2$ respectively represent the magnitude and width of the diffuse scattering signal centered at $q = 0$. The parameters $b_3$, $b_4$ and $b_5$ respectively represent the magnitude, position and width of the charge scattering peak centered at $q = q^*$ and the parameter $b_6$ represents a baseline. Once the peak position and width are fitted to optimal values ($b_4 = q^*$, $b_5 = \omega_c$), we estimate the charge correlation period $p_0 = \frac{2\pi}{q^*}$, interpreted as the average inter-particle distance, and associated correlation length $\lambda_c = \frac{2\pi}{\omega_c}$.

Next, we fit the magnetic signal $|R_M|(q)$, which usually includes a mix of inter-particle magnetic orders. The model function to fit the $|R_M|$ data consists of three pieces: a Gaussian function centered at $q = 0$ with an associated baseline constant to account for the diffuse magnetic scattering signal and for random orientation of nanospins, a Gaussian function accounting for the FM ordering, and a Gaussian function accounting for the AF ordering. By definition, the AF order corresponds to an up/down alternation of neighboring nanospins. If occurring along one given direction (1D) in the material, this alternation theoretically yields a magnetic periodicity that is twice that of the FM order: $p_{AF} = 2\, p_{FM}$. The associated AF scattering signal should then be located near $q \approx q^*/2$. In practice, the fitted location $q^*$ for the charge peak could be used to set the target locations of the FM and the AF peaks, around $q^*$ and $q^*/2$ respectively. However, the XRMS data shows that the AF signal is often extended on a wide range of $q$ values near but not necessarily centered on $\frac{q^*}{2}$. We therefore let the position of both the FM and AF components free to be fitted. The modeled quantity $|R_M|$ is then written as follows:

$$|R_M|\,(\boldsymbol{c}; q) = \frac{c_1}{c_2\sqrt{2\pi}} e^{\frac{-q^2}{2c_2^2}} + \frac{c_3}{c_5\sqrt{2\pi}} e^{\frac{-(q-c_4)^2}{2c_5^2}} + \frac{c_6}{c_8\sqrt{2\pi}} e^{\frac{-(q-c_7)^2}{2c_8^2}} + c_9$$

where $\boldsymbol{c} = (c_1, c_2, \ldots c_9)$ represents a set of fitted coefficients in our model. The coefficient $c_1$ is the relative amplitude of the central diffuse magnetic scattering signal, $c_2$ its width, $c_3$ the relative amplitude of the FM scattering signal, $c_4$ its position (set free to deviate from $q^*$ within $\pm\, \varepsilon$), $c_5$ its width, $c_6$ the relative amplitude of the AF scattering signal, $c_7$ its position (set free to deviate from $\frac{q^*}{2}$ within $\pm\, \varepsilon$), $c_8$ its width and $c_9$ the relative height of an empirical baseline. All the fitting parameters, along with their description, their associated boundaries and the physical symbols associated to them, are summarized in Table 1.

The fitted parameters are used to calculate the following quantities: the FM correlation period $p_{FM} = 2\pi/q_{FM}$ and associated correlation length $\lambda_{FM} = 2\pi/\sigma_{FM}$; the AM correlation period $p_{AF} = 2\pi/q_{AF}$ and associated correlation length $\lambda_{AF} = 2\pi/\sigma_{AF}$. The relative amplitudes $A_{FM}$ and $A_{AF}$ of the FM and AF signal are used to quantify the respective concentrations (in %) of the FM and AF correlations in the NP assembly. In this estimation, we set the FM concentration to be equal to the normalized net magnetization M/Ms since the net magnetization essentially results from FM aligned nanospins.

| Parameter | Description | Boundaries constraints | Symbol for associated physical quantity |
|---|---|---|---|
| $b_1$ | Charge central amplitude | $b_1 \geq 0$ | $A_0$ |
| $b_2$ | Charge central width | $b_2 > 0$ | $\omega_0$ |
| $b_3$ | Charge peak amplitude | $b_3 \geq 0$ | $A_c$ |
| $b_4$ | Charge peak position | $b_4 > 0$ | $q^*$ |
| $b_5$ | Charge peak width | $b_5 > 0$ | $\omega_c$ |
| $b_6$ | Charge baseline | $b_6 \geq 0$ | $B_c$ |
| $c_1$ | Magnetic central amplitude | $c_1 \geq 0$ | $A_1$ |
| $c_2$ | Magnetic central width | $0 < c_2 \leq 1$ | $\sigma_1$ |
| $c_3$ | FM amplitude | $c_3 \geq 0$ | $A_{FM}$ |
| $c_4$ | FM peak position | $q^* \pm \varepsilon$ | $q_{FM}$ |
| $c_5$ | FM width | $0 < c_5 \leq \sigma_{max}$ | $\sigma_{FM}$ |
| $c_6$ | AF amplitude | $c_6 \geq 0$ | $A_{AF}$ |
| $c_7$ | AF peak position | $q^*/2 \pm \varepsilon$ | $q_{AF}$ |
| $c_8$ | AF width | $0 < c_8 \leq \sigma_{max}$ | $\sigma_{AF}$ |
| $c_9$ | Magnetic baseline | $c_9 \geq 0$ | $B_m$ |

Table 1. List of the parameters used in the global fit of the charge and magnetic scattering data.

### B. Synchronous fit of charge and magnetic signals

This global fit was done using the Python library lmfit and its extension of the Levenberg-Marquardt algorithm (LMA) to find the best fit parameters within a restricted domain. The allowed domain for $c_5$ is $q^* \pm \varepsilon$. For M5, $q^* = 1.00$ nm$^{-1}$ and $\varepsilon = 0.08$ nm$^{-1}$ (8% of $q^*$), whereas for M11, $q^* = 0.38$ nm$^{-1}$ and $\varepsilon = 0.06$ nm$^{-1}$ (~16% of $q^*$). Another important constraint is the range for the Gaussian widths $\sigma$. Its upper limit $\sigma_{max}$ corresponds to the lower limit of possible correlation lengths, which we set to the average interparticle distance, namely $\sigma_{max} = 2\pi/p_0$. These constraints, summarized in Table 1, restrict the parameters' domain sufficiently to maintain our mechanistic interpretation of the model.

The data set is comprised of points for each sample at different applied magnetic field values and temperatures. For each field value and temperature, we first fit the charge signal to determine the value of $q^*$ which sets the boundary constraints for the fitting of the $R_m$ signal. The choice of initial parameter values has a large effect on the results of the LMA. In order to ensure that we escape local minima we run the LMA on each data point 300 times with each time new initial parameter guess being a small random deviation from the previous best fit result.

We estimate the standard error in the parameters using the method employed in the lmfit library, namely the square roots of the diagonal elements of the covariance matrix calculated from the model residual Jacobian evaluated at the best fit parameter values. For the error associated with figures of interest derived from parameters. we employed standard error propagation methods to determine the uncertainties in the AF/FM ratio and normalized $A_{FM}$ and $A_{AF}$ values.

A selection of fits for both the charge and magnetic scattering signals at different temperatures and field values for M5 and for M11 is shown in Fig. 5.

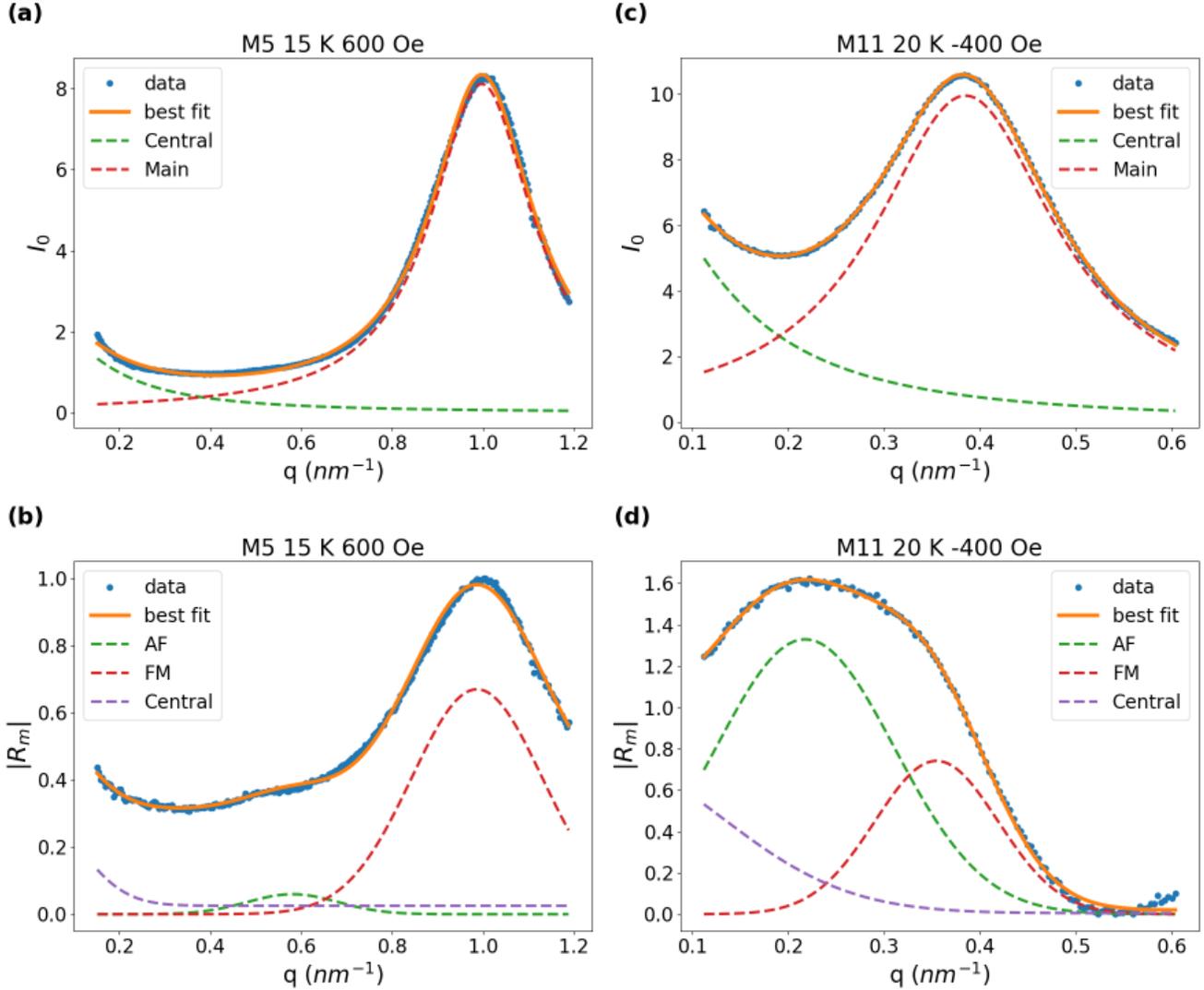

FIG.5: Illustration of the fits carried on the charge scattering intensity $I_0$ and magnetic ratio $R_m$ signals. $I_0$ was fitted using Lorentzian functions and $R_m$ using Gaussian functions. In each case, individual Lorentzian/ Gaussian components are shown in dashed lines. (a) Fit of $I_0$ and (b) fit of $R_m$ for M5 at 15 K and $H = 600$ Oe; (c) Fit of $I_0$ and (b) fit of $R_m$ for M11 at 20 K and $H = -400$ Oe.

## VI. INTERPARTICLE MAGNETIC ORDERING

Below we discuss the results of the fit, from which various physical quantities are extracted, including the periodicity and the correlations lengths for the observed charge and magnetic correlations, as well as their respective strengths, and their evolution throughout the magnetization process at various temperatures for both particle sizes.

### A. Data overview and sorting

An overview of the fitted data is shown in Table 2, where the field values are listed at each temperature, for M5 and for M11. When fitting this data, the parameter relating the AF correlation length showed unrealistic values for some of the datasets. This typically occurred when no apparent AF peak was visible in the magnetic scattering signal $R_M$. In Table 2, the red cells indicate those datasets for which the lower bound on the 95% confidence interval for the AF amplitude (coefficient $c_6$ in the model) reached zero, making the estimate of the associated period and correlation length unreliable. The red cells essentially correspond to data collected at high T for M5 and high field values for M11, where no significant AF component was detected. We chose to not display the fitted AF period nor the associated correlation lengths for these datasets, in subsequent figures 6 and 8.

|  | T (K) | H (Oe) | | | | | | | | | |
|---|---|---|---|---|---|---|---|---|---|---|---|
| M5 | 300 | | | | | 200 | 400 | 600 | 1000 | 2000 | 3000 |
|  | 15 | | | 0 | | | 400 | 600 | 1000 | 2000 | 3000 |
| M11 | 280 | | | | 100 | 200 | 400 | 600 | 1000 | | 3000 |
|  | 20 | -400 | -300 | 0 | 100 | 200 | 400 | 600 | 1000 | | 3000 |

| | No Data | | Lower CI Bound > 0 | | Lower CI Bound includes 0 |

Table 2. Overview of the data collected on M5 and on M11, where the field values are listed for each temperature. Green and red cells correspond to data for which the lower bound on the 95% confidence interval (CI) for the AF amplitude (coefficient $c_6$ in the model) was above zero or included zero, respectively. When $c_6$ is zero, the parameters associated with the peak location and width become unidentifiable and we can make no claims as to their values, therefore we omit these results

## *B. Charge and magnetic periodicity*

The fitted charge periodicity $p_0$ and magnetic periodicities, $p_{FM}$ and $p_{AF}$, are plotted against the applied magnetic field $H$ in Fig.6.

For both M5 and M11, the fitted charge period $p_0$ remains unchanged when $H$ is varied. When averaged over all field values and all temperatures, the average charge period is $p_0 = 6.29 \pm 0.01$ nm for M5 and $p_0 = 16.5 \pm 0.2$ nm for M11. The limited variations (1% or less) across field range [0, 3000 Oe] confirm that the NPs are not physically moving when a magnetic field up to 3000 Oe is applied. No thermal contraction or dilation of the NP assembly is observed, as the interparticle distance remain unchanged throughout cooling and heating. This suggests that, once the solvent is evaporated after deposition on the silicon nitride membrane, the deposited NPs form a rigid monolayer.

The fitted ferromagnetic period $p_{FM}$ remains close to the value of the charge period $p_0$ across the entire field range of field values and temperatures. When averaged across magnetic field $H$ and temperature, the average fitted FM period is $p_{FM} = 6.35 \pm 0.13$ nm (a 2% variation) for M5, and $p_{FM} = 16.8 \pm 0.7$ nm (a 4% variation) for M11. The fact that $p_{FM} \approx p_0$ for both the 5 and 11 nm NPs confirms that each NP remains magnetically monodomain throughout magnetizing and throughout cooling/heating.

When antiferromagnetic signal is present, the fitted AF period $p_{AF}$ lands at a value close to $2p_0$. Namely, on average, $p_{AF} = 12.2 \pm 0.4$ nm for M5 and $p_{AF} = 32.5 \pm 3.9$ nm for M11. $p_{AF}$ shows larger variations (up to $\pm 12\%$) across field values, compared to $p_{FM}$. However, the 95% confidence intervals calculated at each $H$ value goes up to $\Delta p_{AF} = 1.4$ nm for M5 and up to $\Delta p_{AF} = 15$ nm for M11. This suggests that the observed variations on $p_{AF}$ with $H$ may not reflect a real physical change but are mostly caused by uncertainties on locating the AF peak, due to the wideness of the peak, itself caused by the size inhomogeneities, particularly large for M11. Additionally, the fitted $p_{AF}$ values are spread over a range that includes not only $2p_0$ but other remarkable values such as $\sqrt{3}\, p_0$ as visible in Fig. 6.

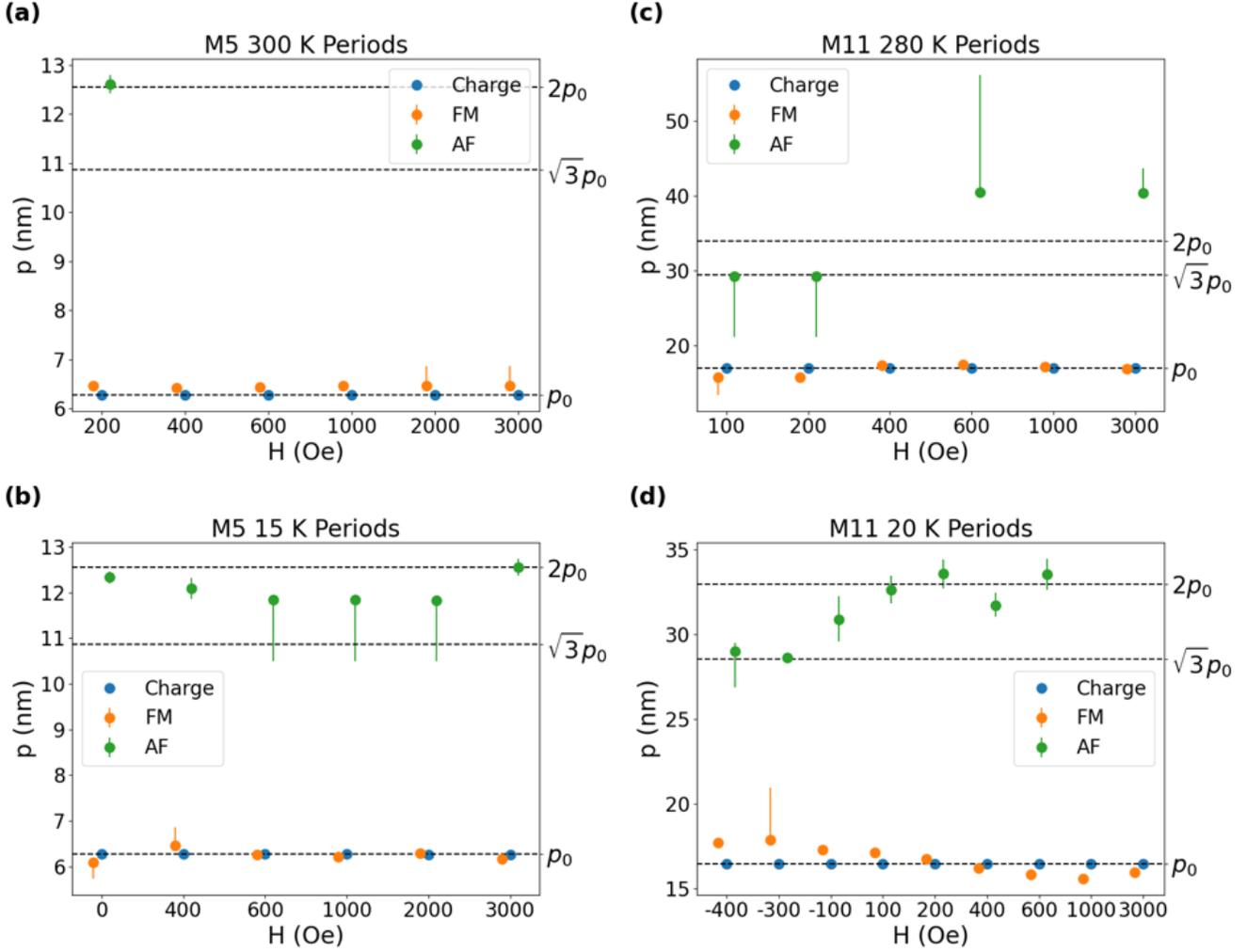

FIG.6 Periodicities extracted from fitting the charge scattering peak in the $I_0$ signal, as well as the FM and AF components in the $R_m$ signal, plotted against the applied magnetic field $H$, for M5 at (a) 300 K and (b) 15 K, and for M11, at (c) 300 K and (d) 20 K. Error bars show the 95% confidence intervals.

When looking more closely at the fitted values for the AF peak location over the entire data set (histogram in Fig.7a), we find that they are distributed around two main values: q*/2 and q*/√3. In real space, this corresponds to periodicities $p_{AF}$ of $2\,p$ and $\sqrt{3}\,p$, respectively. Recalling that the XRMS probe is here sensitive to the out-of-plane component of the magnetization only, the observed magnetic correlations can only relate to arrangements between out-of-plane nanospins. Given that the studied material consists of 2D monolayers of NPs that tend to form hexagonal lattices, one can associate the two observed $p_{AF}$ values to two distinct AF arrangements of out-of-plane nanospins, as illustrated in Fig.7b. One arrangement is a linear type, where the nanospin alternation occurs along one direction, as it would be in a 1D chain of NPs. The AF period in that case is simply twice the nearest neighbor distance $p_{AF} = 2p_0$. Another arrangement is a honeycomb type [51], where the alternation occurs between nanospins located at the six sites of a hexagon. In that case, the AF period is the second next nearest neighbor distance $p_{AF} = 2\cos(30°)\,p_0 = \sqrt{3}\,p_0$. This honeycomb superlattice arrangement partially addresses the well-known spin frustration issue that occurs for AF coupled spins on a triangular (hexagonal) lattice. In the honeycomb superlattice, the six nanospins on the hexagon all satisfy the AF alternation between out-of-plane neighboring spins. Also, it is possible for the center nanospin to be oriented in-plane and not be detected via the XRMS probe. While posing a potentially more pronounced spin frustration issue, the linear AF arrangement is here as widely observed as the honeycomb AF arrangement. This likely occurs, on one hand because of the frequent spatial irregularities in the hexagonal lattice of NPs, resulting sparse islands of close-packed NPs, allowing short-ranged 1D chains of NPs to form at locations. On the other hand, some neighboring nanospins may be oriented in-plane and remain undetected in the XRMS measurement. We emphasize that while the magnetic dipolar couplings favor the AF alignment of spins oriented transversally with respect to the direction of the connecting segment (which is the case for out-of-plane spins), it also favors the FM alignment of spins oriented colinear with the connecting segment (in-plane spins), thus allowing a wide variety of nanospins arrangements, not all visible via XRMS.

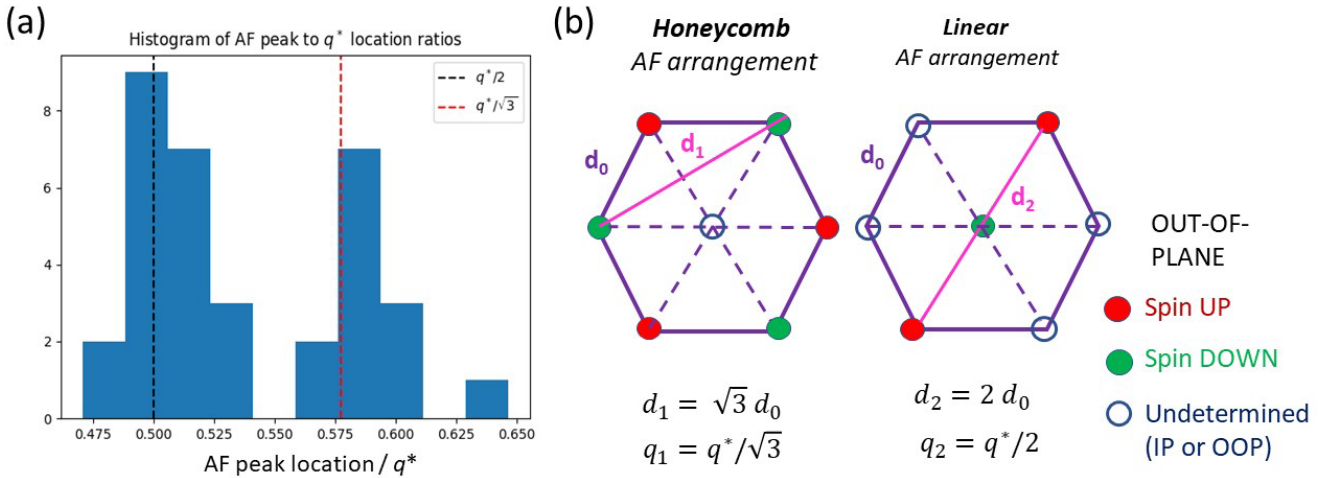

FIG. 7: Visualization of AF arrangements of out-of-plane nanospins in 2D. (a) Histogram of the fitted position $q_{AF}$ of the AF component (coefficient $c_7$) for M5 and M11 combined, clearly showing two dominant values, $q^*/2$ and $q^*/\sqrt{3}$. (b) Illustration of possible AF arrangements of out-of-plane nanospins, on a honeycomb lattice for the $q^*/\sqrt{3}$ component, and in a linear way for the $q^*/2$ component.

### C. Correlation lengths

The fitted correlation lengths for the respective charge, FM and AF orderings are shown in Fig.7. These correlation lengths $\lambda$ are extracted from the fitted peak widths. The correlation length is calculated as follows: when using Gaussian functions, $\lambda = \frac{2\pi}{\sigma}$, and when using Lorentzian function $\lambda = \frac{2\pi}{\omega}\sqrt{2ln2}$ to account for the mathematical difference as well as the fact that $I_0$ represents an intensity and $|R_m|$ represents an amplitude.

The fitted charge correlation length, averaged over all field values and temperatures, is found to be, $\lambda_c = 64.8 \pm 1.1$ nm for M5, which corresponds to about 10.3 NPs being correlated in any direction. For M11, $\lambda_c \sim 53.3 \pm 1.6$ nm, which corresponds, on average, to ~ 3.2 NPs being correlated in any direction. The values for $\lambda_c$ appear independent of the field magnitude and of the temperature. The discrepancy in relative correlation lengths between M5 and M11 is essentially due to particle size inhomogeneity being more pronounced for M11 compared to M5. The higher homogeneity of 5 nm NPs allows the particles to self-assemble and structurally correlate over longer ranges, compared the more inhomogeneous 11 nm NPs.

The fitted FM correlation length $\lambda_{FM}$ remains relatively stable across field values for both M5 and M11. Like for the charge correlations, the FM order appears to extend over a larger number of NPs in the case of the 5 nm NPs compared to the case of 11 nm NPs. For M5, the average FM correlation length at room temperature is $\lambda_{FM} =$ $60 \pm 12$ nm, which corresponds to about 9.6 NPs being ferromagnetically correlated in any direction. For M11, $\lambda_{FM} = 70 \pm 20$ nm, corresponding to about 4.2 NPs being ferromagnetically correlated in any direction.

These values do not significantly change when cooling down below $T_B$.

The fitted value for $\lambda_{AF}$ shows large (up to 100 %) uncertainties at certain field values, so the observed variations of $\lambda_{AF}$ with field $H$ is not meaningful.

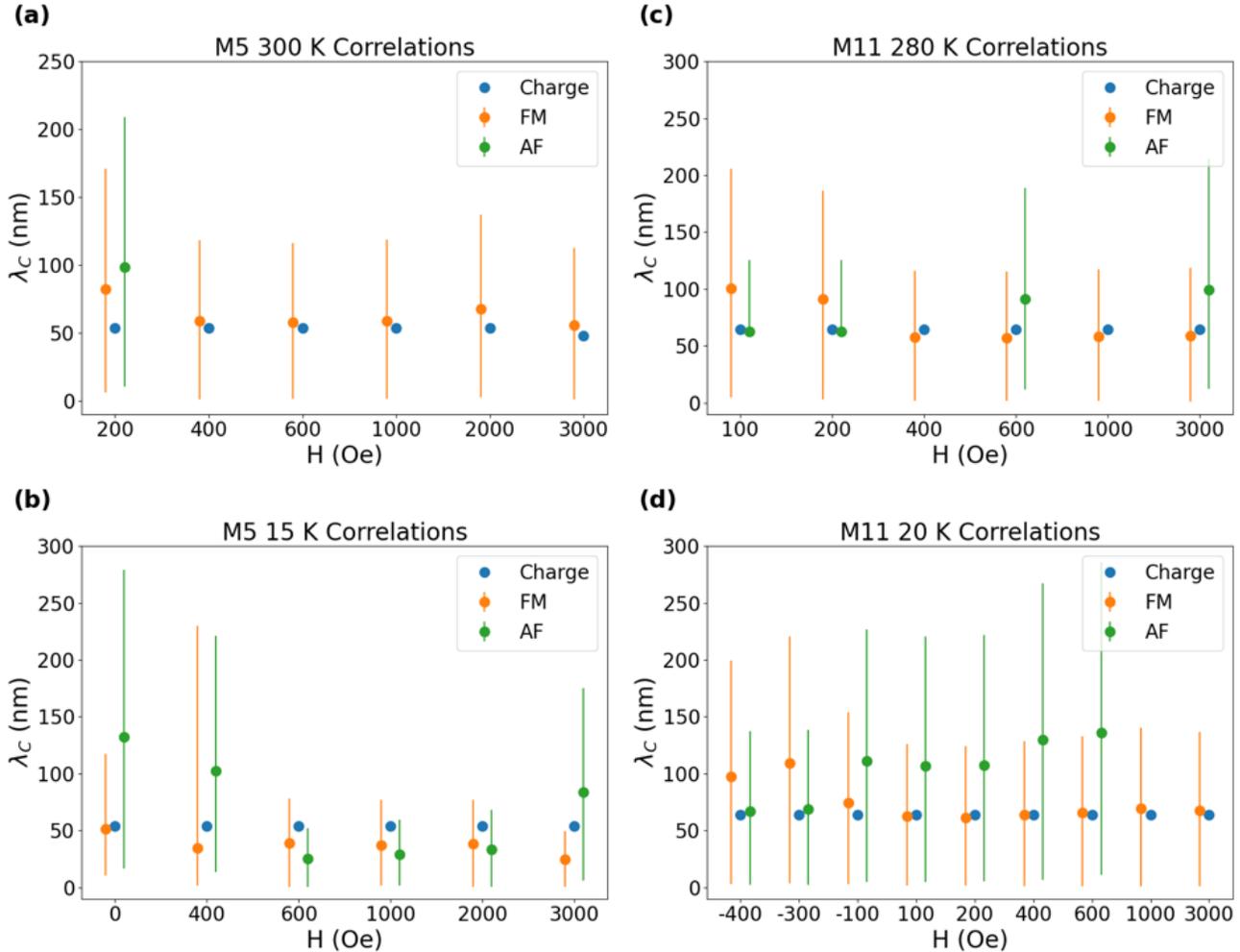

FIG.8 Correlation lengths extracted from fitting the charge scattering peak in the $I_0$ signal, as well as the FM and AF components in the $R_m$ signal, plotted against the applied magnetic field $H$, for M5 at (a) 300 K and (b) 15 K, and for M11, at (c) 300 K and (d) 20 K. Error bars show the 95% confidence intervals.

### D. Magnetic ordering vs. applied magnetic field

The relative amplitudes associated to the fitted FM and AF components (coefficient $c_3$ and $c_6$ in the model) and their evolution throughout the magnetization process are shown in Fig.8. For comparison purposes, the FM amplitude $c_{FM}$ is set to the value of the normalized net magnetization $M/Ms$ that was measured via VSM (except near the coercive point where, by definition, $M = 0$ and instead the relative magnitude of the magnetic scattering signal $|R_M|$ is used). This normalized concentration $c_{FM}$ is interpreted as the portion of the NP assembly where nanospins align ferromagnetically out-of-plane. The normalized AF concentration $c_{AF}$ is accordingly set as $c_{AF} = (c_6 / c_3) M/Ms$ and corresponds to the portion of the NP assembly where nanospins align antiferromagnetically out-of-plane. The sum $c_{FM} + c_{AF}$ should not exceed 1 or 100%. The remaining portion, $1 - c_{FM} - c_{AF}$, corresponds to the portion of the material that is neither FM or AF correlated out-of-plane, and which we identify as magnetic randomness. These random nanospins may be aligned either out-of-plane or in-plane. The fitted values for these various coefficients for M5 and M11, respectively are summarized in Table 3.

Both M5 and M11 display the following trends at room temperature (above $T_B$): 1) the FM concentration decreases as the field decreases from 3000 Oe down to zero according to the respective magnetization curves; 2) the AF concentration at room temperature remains relatively small (less than 2.3% for M5 and less than 9.2% for M11) across all field values; 3) as the field is decreased, the magnetic randomness increases and gradually takes over, reaching up to 93 % in the case of M5 and 78 % in the case of M11. This confirms the predominant superparamagnetic nature of the NP assemblies above $T_B$.

*E. Dependence on temperature*

Significant changes in the magnetic behavior are observed when cooling the material from above to below $T_B$. While magnetic randomness remains predominant, a non-negligible AF ordering component emerges at low field values when below $T_B$.

For M5, we observe the following: at 300 K, the randomness concentration starts above 42% at 3000 Oe and increases up to 93% when $H$ is decreased to zero (remanence). Throughout the entire field range, no significant AF correlations are measured (AF concentration remains less than 2.3%). At 15 K below $T_B$, M5 exhibits a similar trend, except the randomness concentration at 3000 Oe is not as high, down to 14%. As the field decreases to zero, the randomness increases up to 81% at remanence, where small but non-negligible AF correlations emerge, covering up to 12%. While the magnetization data for M5 indicates no sign of significant hysteresis neither at 300 K nor at 15 K, the XRMS data reveals the presence of non-zero AF correlations between out-of-plane nanopsins near remanence. The XRMS data also allows to quantify the amount of these AF correlations and see that it sensibly increases (from ~ 2 to 10 %) when cooling the material below $T_B$.

In the case of M11, AF correlations between out-of-plane nanospins are also observed, and they appear to be even stronger than for M5. Above $T_B$, the measured AF concentration reaches up to 9.2 % at 300 K. When the material is cooled down to 20 K, below $T_B$ where significant hysteresis is observed ($H_c$ = 300 Oe), and when the field is decreased near the coercive point, the AF concentration significantly increases, covering up to 35 % of the NP assembly, with a confidence interval for this parameter ranging from 29% to as high as 48 %. Here again, not only the XRMS data corroborates the VSM data it shows hysteresis but allows to identify the nature and extent of the magnetic interactions between out-of-plane nanopsins at the coercive point.

While discussing temperature dependence, one should account for possible time averaging effects due to the inherent magnetic fluctuation of the nanospins in the NP assemblies. Indeed, in the SPM phase, and in the absence of magnetic field, the nanospins supposedly randomly fluctuate at a time scale comparable to, or even faster than the measurement time, preventing the observation of any magnetic ordering. However, when cooling the NP assemblies below their blocking temperature, the dynamics of magnetic fluctuation drastically slows down, and, at low T well below $T_B$, the nanospin distribution may appear almost static at the timescale of the measurement. It is then possible to observe instantaneous magnetic orderings occurring between the nanospins. In a separate study (not published yet), we have measured characteristic fluctuation times ranging from ~ 200 sec at 200 K up to ~ 5000 sec at 15 K for M5 (for which $T_B$ ~ 25 K), and from ~4000 sec at 250 K up to ~50,000 sec at 100 K for M11 with $T_B$ ~ 170 K). Based on this data, we can comfortably assume that during the XRMS collection time (~200 sec), the NP assemblies are mostly static, with little effect from magnetic fluctuations on the observed magnetic correlations.

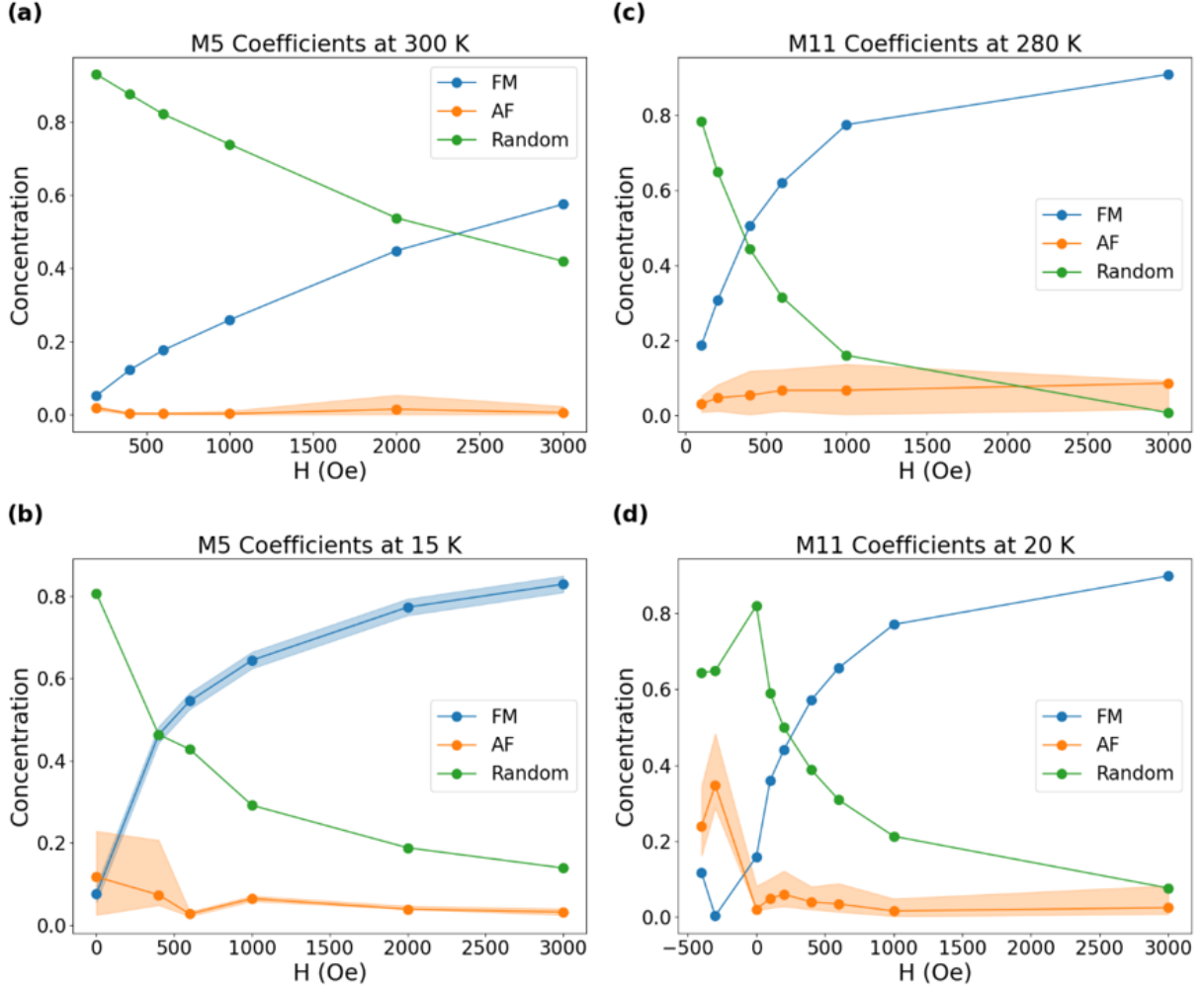

FIG.9 Concentrations for the FM and AF correlations as a function of field H, above and below $T_B$. (a) For M5 at 300 K; (b) for M5 at 15 K; (c) for M11 at 280 K; (d) for M11 at 20 K.

## F. Dependence on particle size

The differences in the magnetic behavior of M5 vs. M11, suggest a possible dependence on NP size. To identify a possible trend, we have plotted, in Fig.9, the relative ratio $c_{AF}/c_{FM}$ of the AF and FM components as a function of $H$. For both M5 and M11, and at all temperatures, this ratio remains relatively small (less than 0.1) in most of the field range. Only when the field approaches zero (or the coercive point in case hysteresis exists), the AF/FM ratio starts increasing drastically. At that stage, a noticeable difference is observed between M5 and M11: for M5 the AF/FM ratio goes up to ~ 1.6 with a 95% confidence interval ranging from 0.4 up to 3.1, whereas for M11 the AF/FM ratio goes up to ~ 3.8 with a 95% confidence interval ranging from 3.2 up to 5.4. This suggests that the NP particle size may affect the strength and extent of AF correlations when present: the larger 11 nm NPs show a higher AF concentration compared to the 5 nm NPs. This effect is likely caused by the drastic increase in the magnitude of the nanospin carried by individual NP when the NP size increases. In theory, if ignoring surface effects, this increase goes cubically with the particle size. We estimated the nanospin moment for the 5 nm NPs to be $M_n \sim 1400\ \mu_B$, whereas for the 11 nm NPs, $M_n \sim 14{,}000\ \mu_B$, that is about 10 times stronger. The dipolar magnetic interaction between neighboring nanospins, which goes as $E \sim \frac{M_n^2}{p^3}$, increases quadratically with $M_n$. Assuming the interparticle distance $p$ is fixed, this dipolar interaction is amplified by nearly a factor 100 when increasing the NP size from 5 to 11 nm. In reality, the average interparticle distance $p$ also increases, from 6.3 nm to 16.5 nm, when going from M5 to M11. However, even when accounting for the larger spacing, the dipolar interaction still increases by about a factor 5 when going from 5 nm to 11 nm NPs. Consequently, the AF correlations are stronger for the larger 11 nm NPs, compared to the 5 nm NPs.

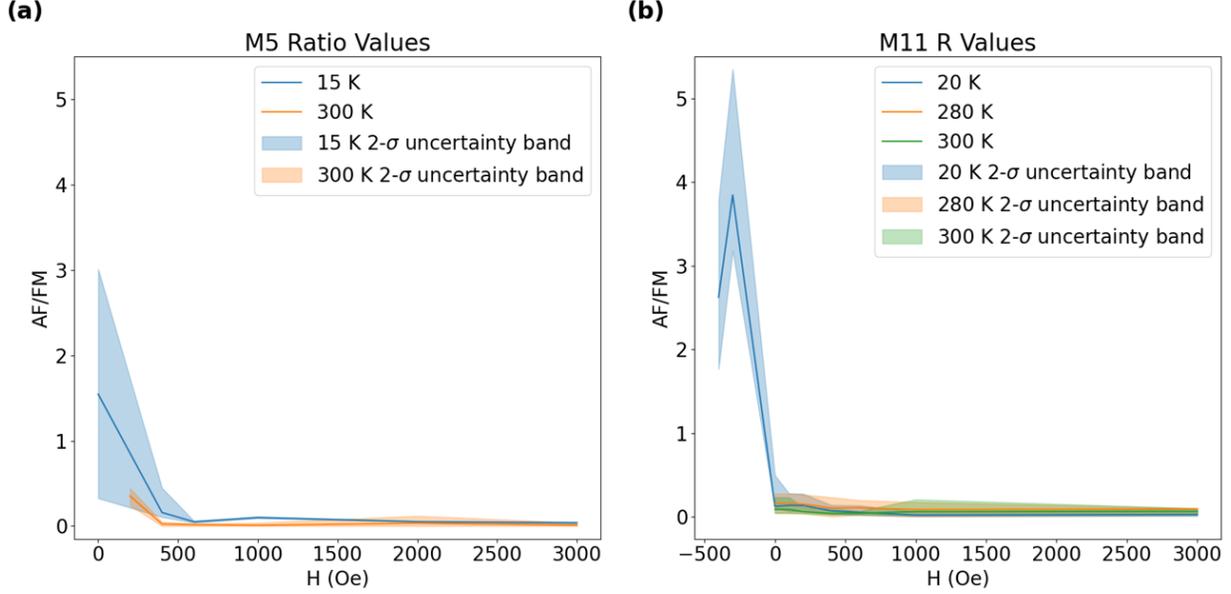

FIG.10 Relative AF/FM ratio vs. *H* field at high and low T for (a) M5 and (b) M11.

|  | M5 (5 nm) |  | M11 (11 nm) |  |
| ---: | :---: | :---: | :---: | :---: |
| $T_B$ | 28 K | | ~170 K | |
| Average charge period $p_0$ | 6.29 ± 0.01 nm | | 16.5 ± 0.2 nm | |
| Aver. charge correlation $\lambda_c$ | 64.8 ± 1.1 nm | | 53.3 ± 1.6 nm | |
| Aver. FM period $p_{FM}$ | 6.35 ± 0.13 nm | | 16.8 ± 0.7 nm | |
| Aver. FM correlation $\lambda_{FM}$ | 60 ± 12 nm | | 70 ± 20 nm | |
| Aver. AF period $p_{AF}$ | 12.2 ± 0.4 nm | | 32.5 ± 3.9 nm | |
|  | 300 K | 15 K | 280 K | 20 K |
| $H_c$ | 0 Oe | 0 Oe | 0 Oe | - 300 Oe |
| $M/M_s$ @ 3000 Oe | 0.58 | 0.69 | 0.91 | 0.89 |
| Randomness @ 3000 Oe | 42% | 14% | 0.7% | 7.6% |
| Max Randomness | 93% | 81% | 78% | 82% |
| Min AF concentration $A_{AF}$ | 0.2% | 3.2% | 2.3% | 1.6% |
| Max AF concentration $A_{AF}$ | 2.3% | 12% | 9.2% | 35% |
| Max ($A_{AF}$ / $A_{FM}$) average ratio | 0.35 | 1.6 | 0.16 | 3.8 |

Table 3. Measured and fitted values of selected parameters for M5 and M11, including blocking temperature ($T_B$), coercive field ($H_c$), normalized magnetization ($M/M_s$) at 3000 Oe, fitted periods and correlation lengths for the charge, FM and AF contributions, fitted concentrations for randomness and AF order, fitted AF/FM concentration ratio.

# VII. CONCLUSION

Our x-ray resonant magnetic scattering (XRMS) study of monolayers of magnetite NPs has unveiled nanoscale interparticle magnetic correlations within 2D assemblies of nanospins, that are unseen via other conventional magnetometry techniques, and difficult to access via SANS. The XRMS data is particularly instructive when the net magnetization is near zero (which occurs near zero field when in the SPM state, or near the coercive point when hysteresis exists), and there is question whether the nanospins are randomly oriented or are magnetically correlated.

If the magnetometry data shows no hysteresis, it is generally assumed that, at zero magnetization, the nanospins are magnetically uncorrelated and magnetic randomness is expected. However, our XRMS data revealed that even in the absence of visible hysteresis, nanoscale magnetic interparticle correlations between out-of-plane nanospins may exist. Even in a small extent (in the range of few %), these correlations, invisible via standard magnetometry, can be detected and quantified via XRMS.

If, on the other hand, hysteresis occurs in the magnetization loop, the hysteresis is commonly attributed to various magnetic anisotropies within the NPs, but little is known about possible interparticle magnetic interactions. The XRMS technique allows to visualize the nanoscale magnetic correlations at any point of the magnetization loop, and thus reveals the nature of the reversal processes through the coercive point. Our XRMS data shows that, when applying out-of-plane magnetic field, the hysteresis is accompanied by significant interparticle magnetic correlations and that, at the coercive point, extended AF correlations occurs between out-of-plane nanospins.

For both the 5 nm and 11 nm NPs, we found that, when above their respective blocking temperatures $T_B$, magnetic randomness, where the nanospins may be pointing either out-of-plane or in-plane (non-detected), largely dominates when the magnetic field is released back to zero, confirming the SPM nature of the NP assemblies. However, when these NP assemblies are cooled down below $T_B$, significant out-of-plane interparticle AF correlations emerge when the magnetic field is released. For the smaller 5 nm NPs, even though no magnetic hysteresis was observed at 15 K, AF ordering covering up to 12% of the assembly was detected. For the larger 11 nm NPs that do exhibit visible hysteresis below $T_B$, more extended AF correlations, covering up to 48% of the assembly were detected at 20 K. These largely extended out-of-plane interparticle magnetic correlations, which we may describe as super-antiferromagnetic (SAF), can be physically explained by magnetic dipole couplings, that favors the antiparallel alignment of neighboring dipoles in the direction transverse to the connecting segment.

Based on these findings, useful directions to investigate in the future include further studying the emergence of interparticle magnetic couplings in assemblies of magnetic NPs and their dependence on NP size and shape, as well as on NP concentration in the assembly. This tool can be useful for medical applications, such as hyperthermia, drug delivery and gene delivery, where functionalized encapsulated magnetic NPs are manipulated via an external magnetic and their collective response strongly depends on the presence of interparticle magnetic couplings. If interparticle magnetic couplings are undesired in these applications, one could learn to eliminate or control them by adjusting the NP size, structure and environment. In conclusion, nanoscale magnetic correlations in NP materials are still largely unknown and to be unveiled. The XRMS technique allows the detection of interparticle magnetic correlations that cannot be seen via standard magnetometry techniques and provides a way to potentially control and tailor these magnetic correlations for various applications.

**Acknowledgments**

We thank the College of Physical and Mathematical Science at BYU for funding graduate and undergraduate students involved in this research, as well as the TEM and VSM instruments. The XRMS data was collected at SSRL at SLAC National Laboratory, supported by the U.S. Department of Energy, Office of Science User Facility.